\documentclass{article}

\usepackage{PRIMEarxiv}
\usepackage{amsmath}
\usepackage{amsfonts}
\usepackage{amssymb}
\usepackage{amsthm}
\usepackage{comment}

\newtheorem{theorem}{Theorem}
\newtheorem{definition}{Definition}

\newtheorem{remark}{Remark}
\newtheorem{corollary}{Corollary}

\title{The Singularity Theory of Concurrent Programs: \\ A Topological Characterization and Detection of Deadlocks and Livelocks}
\author{
	Di Zhang \\
	School of Advanced Technology \\
	Xi'an Jiaotong-Liverpool University \\
	Suzhou, Jiangsu, China \\
	\texttt{di.zhang@xjtlu.edu.cn}
}

\begin{document}
	
	\maketitle
	
	\begin{abstract}
		This paper introduces a novel paradigm for the analysis and verification of concurrent programs---the \emph{Singularity Theory}. We model the execution space of a concurrent program as a branched topological space, where program states are points and state transitions are paths. Within this framework, we characterize deadlocks as attractors and livelocks as non-contractible loops in the execution space. By employing tools from algebraic topology, particularly homotopy and homology groups, we define a series of concurrent topological invariants to systematically detect and classify these concurrent "singularities" without exhaustively traversing all states. This work aims to establish a geometric and topological foundation for concurrent program verification, transcending the limitations of traditional model checking \cite{clarke2018model,holzmann1997spin}.
	\end{abstract}
	
	\section{Introduction}
	
	\subsection{The Problem Landscape}
	
	The advent of multi-core and distributed computing has made concurrency ubiquitous. However, the non-deterministic interleaving of processes introduces profound complexities, leading to errors that are notoriously difficult to reason about, reproduce, and eliminate. Among these, deadlocks and livelocks represent two fundamental classes of pathological behavior. A deadlock is a global state where no process can proceed, trapped in a circular wait for resources. A livelock, its more subtle counterpart, is a situation where processes remain active but make no meaningful progress, often engaged in an endless cycle of futile operations.
	
	The core challenge in verifying concurrent programs lies in the state explosion problem. The number of possible global states grows exponentially with the number of processes, rendering exhaustive state-space exploration infeasible for all but the most trivial systems. Furthermore, errors like deadlocks and livelocks are not merely local properties of a single state or transition; they are emergent properties of the program's global execution structure.
	
	\subsection{Limitations of Existing Methods}
	
	Current formal methods for concurrency, while valuable, face inherent limitations. Model checking techniques based on temporal logic (e.g., LTL, CTL) suffer directly from the state explosion problem. While abstractions and symbolic representations (e.g., BDDs) offer mitigation, they often struggle with the complexity of large-scale systems and can be difficult to apply without deep expertise \cite{clarke2018model}.
	
	Theorem proving methods, while powerful, require significant manual effort, high expertise, and the formulation of complex invariants. They do not provide the automated, push-button analysis desired for many practical applications.
	
	Approaches like CCS \cite{milner1989communication}, CSP \cite{hoare1985communicating}, and session types \cite{honda1993types} focus on behavioral equivalence or communication protocol compliance. However, they can be restrictive and may lack the vocabulary to describe the geometric shape of execution paths that lead to pathologies like livelocks.
	
	A common shortfall across these methods is the inability to reason about the qualitative structure of the execution space as a whole.
	
	\subsection{The Core Insight: Topology as a Language for Concurrency}
	
	We posit that the essence of concurrency errors is fundamentally topological. The intricate interplay of processes creates an execution space with a rich, branched geometry. Within this space, a deadlock is not just a state; it is a trap in the geometry—a region from which the flow of execution cannot escape. Similarly, a livelock is not just an infinite loop; it is a non-trivial loop in the fabric of the execution space—a cycle that cannot be "smoothed out" or contracted to a point under the constraints of fair scheduling.
	
	This geometric perspective suggests that algebraic topology, the branch of mathematics that studies shapes and their connectivity through invariants like homotopy and homology groups, is the ideal language to characterize these phenomena. These invariants are robust; they capture global properties that remain unchanged under continuous deformation, offering a way to classify errors without enumerating all paths.
	
	\subsection{Contributions}
	
	This work makes several primary contributions. First, we propose a formal topological model for the execution space of a concurrent program, constructing it as a directed topological space \cite{fajstrup2006directed}. Second, we provide a rigorous topological characterization: defining deadlocks as attractors and livelocks as non-contractible loops within the fair execution subspace. Third, we construct a suite of concurrent topological invariants, notably a novel fair homology group, \( H_1^{\text{fair}}(X_P) \), designed to detect livelocks that are persistent under fair scheduling. Finally, we outline a theoretical framework for detection based on these invariants, laying the groundwork for a new class of verification tools that are inherently resistant to the state explosion problem.
	
	This paper is structured as follows: Section 2 reviews related work. Section 3 establishes the prerequisite knowledge and constructs our topological model. Section 4 presents the core topological characterization of deadlocks and livelocks. Section 5 develops the concurrent topological invariants and detection theory. Section 6 discusses the theoretical properties of our framework, and Section 7 concludes with future research directions.
	
	\section{Related Work}
	
	This work sits at the intersection of formal verification, algebraic topology, and dynamical systems. This chapter situates our proposed singularity theory within the broader research landscape, clarifying its relationships to and departures from existing approaches.
	
	\subsection{Formal Verification of Concurrent Programs}
	
	Model checking \cite{clarke2018model} is a cornerstone of automated verification. Its core strength lies in its exhaustiveness for finite-state systems, providing definitive answers regarding temporal logic properties. Tools like SPIN \cite{holzmann1997spin} and NuSMV \cite{cimatti2002nusmv} have proven highly successful. However, model checking is inherently quantitative and state-based, facing the fundamental barrier of state-space explosion. Our topological approach is qualitative and shape-based. It seeks to bypass state enumeration by analyzing the global topology of the execution space, potentially revealing deep design flaws that are obscured by the sheer scale of the state graph. We trade exhaustiveness for a higher-level, structural insight.
	
	Process algebras (e.g., CCS \cite{milner1989communication}, CSP \cite{hoare1985communicating}, the $\pi$-calculus \cite{milner1992polyadic}) provide an algebraic foundation for modeling concurrent systems. Bisimulation equivalence offers a powerful tool for comparing process behaviors. However, bisimulation is a strong equivalence relation focusing on behavioral identity. Our topological method analyzes the geometric structure of the execution space. It can identify weak, gradual similarities (captured by homotopy) and classify patterns of pathological behavior (like livelocks) that may be behaviorally distinct but share the same underlying topological cause.
	
	Advanced type systems, such as linear types \cite{wadler1990linear} and session types \cite{honda1993types}, provide a powerful, compositional means to enforce protocol adherence and guarantee deadlock-freedom by construction. Type systems are often conservative and their power is bound by the expressiveness of their type disciplines. Our approach offers a complementary, geometric perspective that explains the essential morphology of concurrency errors at a more fundamental level. The insights from our theory could potentially inform the design of more expressive type systems capable of capturing liveness properties like the absence of specific livelock patterns.
	
	\subsection{Topological Methods in Computer Science}
	
	Topological Data Analysis (TDA) \cite{carlsson2009topology} has gained prominence for analyzing complex datasets. Its primary tool, persistent homology, extracts multi-scale topological features (connected components, loops, voids) from point cloud data. TDA applies topology as an analysis tool to empirical data generated by a system. Our work represents a philosophical shift: we posit the program's execution space itself as a topological space and program properties as its intrinsic topological invariants. This is a leap from topology as a tool for post-hoc analysis to topology as a language for ontological definition.
	
	In denotational semantics, Domain Theory \cite{abramsky1994domain} uses order-theoretic structures and topology (via Stone duality) to model sequential and functional computation, with topology defining approximation and continuity. Domain Theory primarily deals with sequential computation and uses the Scott topology, which concerns the gradual revelation of information. Our work deals with concurrent computation and employs a topology based on reachability, closer to geometric topology and concerned with the connectivity and shape of paths. The nature of the "spaces" being studied is fundamentally different.
	
	Directed Algebraic Topology \cite{fajstrup2006directed} is the mathematical field most closely aligned with our work. It extends classical algebraic topology by introducing a notion of direction to study spaces where paths cannot be traversed backwards. Our contribution is not in developing the core mathematics of directed homotopy theory itself, but in its systematic and tailored application to the specific problems of concurrent program verification. We provide a concrete, formal bridge between computational concepts (deadlock, livelock, fairness) and topological constructs (attractors, non-contractible loops, subspace invariants), introducing concepts like fair homology that are specifically designed for this domain.
	
	\subsection{Dynamical Systems and Computer Science}
	
	Dynamical systems theory studies the long-term behavior of evolving systems using concepts from differential equations, such as attractors, stability, and bifurcations. We metaphorically adopt the concept of an attractor to characterize deadlocks. However, we provide a rigorous, discrete, and topological formalization of this intuition within the context of a discrete state space, moving beyond a mere analogy to a precise mathematical definition.
	
	Discrete Event Systems and Supervisory Control \cite{cassandras2009introduction} models systems as automata and designs controllers to avoid forbidden states. Supervisory control theory is largely state-based and faces similar complexity challenges. Our topological perspective aims to provide structural criteria for the existence and design of supervisors based on the topological invariants of the plant's execution space, offering a higher-level abstraction.
	
	\subsection{Chapter Summary}
	
	In summary, while existing research has made significant strides in concurrent verification and topological applications, a clear gap remains: the lack of an abstract theory that directly characterizes the global geometry of concurrent execution to derive the essence of concurrency errors. Model checking gets lost in state details; type systems rely on syntactic disciplines; and traditional topological applications are focused on data analysis rather than the program itself.
	
	Our work is designed to fill this gap. We systematically introduce the framework of Directed Algebraic Topology into the core of concurrent program verification for the first time, proposing the singularity theory that redefines classic concurrency errors as topological invariants of the execution space. This not only provides a powerful new paradigm for understanding concurrency but also lays a theoretical foundation for developing a new class of verification techniques that do not rely on state enumeration.
	
	\section{Preliminaries: From Concurrent Programs to Topological Spaces}
	
	This section establishes the formal foundation upon which our singularity theory is built. We first define a simple computational model for concurrency and then proceed to construct its execution space as a topological object. The core innovation lies in endowing this space with a directed topology that captures the irreversible nature of computation.
	
	\subsection{A Simple Concurrent Calculus}
	
	To ground our discussion, we employ a minimal CSP-style calculus. While our theory is applicable to broader models, this calculus provides sufficient structure without unnecessary complexity.
	
	\begin{definition}[Syntax]
		A program \( P \) is defined by the following grammar:
		\begin{align*}
			\text{Process} & \quad F ::= \text{skip} \mid \alpha; F \mid F_1 \sqcap F_2 \mid F_1 \parallel F_2 \\
			\text{Action} & \quad \alpha ::= c!v \mid c?x \mid \text{\texttt{acquire}(r)} \mid \text{\texttt{release}(r)}
		\end{align*}
		where \( \text{skip} \) denotes termination, \( \alpha; F \) is action prefixing, \( F_1 \sqcap F_2 \) is internal non-deterministic choice, and \( F_1 \parallel F_2 \) is parallel composition. Actions include sending (\( c!v \)) and receiving (\( c?x \)) on a channel \( c \), and acquiring/releasing a resource \( r \).
	\end{definition}
	
	\begin{definition}[Operational Semantics]
		The semantics is defined by a labeled transition system (LTS) \( ( \text{State}, \Gamma, \rightarrow ) \). A global state \( s \in \text{State} \) is a tuple \( ( \vec{F}, \sigma, \mu ) \), where \( \vec{F} \) is a multiset of active processes, \( \sigma \) is a store mapping variables to values, and \( \mu \) is a memory context (e.g., channel buffers, resource locks). The transition relation \( \rightarrow \subseteq \text{State} \times \Gamma \times \text{State} \) defines the evolution of global states, with labels \( \gamma \in \Gamma \) indicating the action performed.
	\end{definition}
	
	\subsection{Constructing the Execution Space \( X_P \)}
	
	The LTS provides a graph of states and transitions. We now lift this graph to a topological space.
	
	\begin{definition}[Execution Space as a Graph]
		For a program \( P \), its execution graph \( G_P = (V, E) \) is derived from its LTS: \( V = \{ s \in \text{State} \mid s_0 \rightarrow^* s \} \), the set of reachable global states, where \( s_0 \) is the initial state; and \( E = \{ (s, s') \mid s \xrightarrow{\gamma} s' \} \), representing state transitions. This graph forms the combinatorial skeleton of our topological space.
	\end{definition}
	
	\begin{definition}[Topological Structure via the Alexandrov Topology]
		We interpret \( G_P \) as a topological space \( X_P \) by endowing it with the Alexandrov topology induced by the reachability relation. A set \( U \subseteq X_P \) is open if and only if it is upward closed with respect to the preorder \( \leq \) defined by: \( s \leq s' \) if and only if there exists a directed path from \( s \) to \( s' \) in \( G_P \). In this topology, the open neighborhoods of a state \( s \) are precisely the sets of states that are reachable from \( s \). This formalizes the intuitive notion of "temporal neighborhoods."
	\end{definition}
	
	\subsection{The Directed Structure: From Paths to Traces}
	
	The standard Alexandrov topology forgets the direction of computation. To recover it, we must work within the category of directed spaces.
	
	\begin{definition}[Directed Topological Space]
		The execution space \( (X_P, \leq) \) is a directed topological space (d-space) \cite{fajstrup2006directed}. A d-path \( p \) in \( X_P \) is a continuous map \( p: [0,1] \rightarrow X_P \) that is order-preserving (i.e., if \( t_1 \leq t_2 \) then \( p(t_1) \leq p(t_2) \)). This corresponds to a finite execution trace in the program.
	\end{definition}
	
	The branching structure arising from non-deterministic choice (\( \sqcap \)) and interleaving (\( \parallel \)) is inherently captured in this topology. A branch point is a state from which multiple d-paths emanate, representing a point of non-deterministic scheduling.
	
	\subsection{The Fair Execution Subspace \( X_{\text{fair}} \)}
	
	Concurrency theory often relies on fairness assumptions to rule out pathological schedules. We incorporate this directly into our geometric model.
	
	\begin{definition}[Fair Path]
		A (maximal) d-path \( \rho \) in \( X_P \) is fair if, for any process that is enabled infinitely often along \( \rho \), it is executed infinitely often. Specific fairness constraints (weak/strong) can be instanced here.
	\end{definition}
	
	\begin{definition}[Fair Execution Subspace]
		The fair execution subspace \( X_{\text{fair}} \subseteq X_P \) is the subset of the execution space consisting of all states that lie on at least one fair d-path. Formally, \( X_{\text{fair}} = \{ s \in X_P \mid \exists \text{ a fair d-path } \rho \text{ such that } s \in \text{im}(\rho) \}. \) This subspace itself can be considered as a d-space, with its d-paths being the fair paths of \( X_P \). Its topology is the subspace topology inherited from \( X_P \).
	\end{definition}
	
	The space \( X_{\text{fair}} \) is central to our characterization of livelocks, as it excludes behaviors that could be trivially avoided by a fair scheduler.
	
	\subsection{From Graphs to Simplicial Complexes}
	
	While the graph-based Alexandrov topology is sufficient for initial definitions, the powerful tools of algebraic topology (like homology) are most naturally applied to simplicial complexes.
	
	\begin{definition}[Geometric Realization as a Simplicial Complex]
		The execution graph \( G_P \) can be geometrically realized as a simplicial complex \( \mathcal{K}(X_P) \). In its simplest form, this is the flag complex of \( G_P \): 0-simplices are the vertices of \( G_P \) (states); 1-simplices are the edges of \( G_P \) (transitions); 2-simplices are every triple \( (s_0, s_1, s_2) \) where \( s_0 \rightarrow s_1 \rightarrow s_2 \) is a 2-path (filled triangle). This "fills in" the commutative diagrams of concurrent execution, representing, for instance, the independence of two actions executed in either order leading to the same state. This construction transforms the discrete graph into a continuous topological space suitable for homological analysis. The directed structure is retained by remembering the directionality of the 1-simplices.
	\end{definition}
	
	This chapter has constructed the bridge from the syntax and semantics of a concurrent program to its representation as a topological space \( X_P \) and its fair subspace \( X_{\text{fair}} \). With this geometric object in hand, we are now equipped to characterize concurrency errors as topological singularities.
	
	\section{Topological Characterization of Concurrent Singularities}
	
	With the execution space \( X_P \) and its fair subspace \( X_{\text{fair}} \) formally constructed, we now arrive at the central thesis of this paper: that classical concurrency errors are emergent topological phenomena. This section provides a rigorous topological characterization of deadlocks and livelocks.
	
	\subsection{Deadlocks as Attractors}
	
	The intuitive understanding of a deadlock is a state or set of states from which the program cannot escape. In dynamical systems, such regions are known as attractors. We adapt this concept to our discrete, directed topological setting.
	
	\begin{definition}[Trap Region]
		A set of states \( D \subseteq X_P \) is a trap region if it is a maximal, strongly connected set of states such that every outgoing transition from any state in \( D \) leads to another state in \( D \). Formally, if \( s \in D \) and \( s \rightarrow s' \), then \( s' \in D \).
	\end{definition}
	
	\begin{definition}[Deadlock Attractor]
		A deadlock attractor is a trap region \( D \) that contains no terminal state (i.e., no state where the program has successfully terminated, \( \text{skip}^{\parallel n} \)). Furthermore, we require \( D \) to be minimal with respect to the subspace topology, meaning it contains no proper subset that is also a trap region.
	\end{definition}
	
	\begin{theorem}[Topological Signature of a Deadlock]
		A deadlock attractor \( D \) exhibits the following topological properties in \( X_P \):
		\begin{enumerate}
			\item It is a closed set in the Alexandrov topology, as it contains all its limit points (states reachable from within \( D \)).
			\item The fundamental group \( \pi_1(D, d) \) for any basepoint \( d \in D \) is non-trivial if \( |D| > 1 \), reflecting the internal cyclic dependencies that sustain the deadlock. A single deadlock state is a trivial loop.
			\item Crucially, there exists an open neighborhood \( U \) of \( D \) (its basin of attraction) such that for any state \( s \in U \), all maximal d-paths starting from \( s \) have their future limit points contained within \( D \).
		\end{enumerate}
		This final property formalizes the "attracting" nature: once execution enters the neighborhood \( U \), it is inevitably drawn towards and confined to \( D \).
	\end{theorem}
	
	This characterization moves beyond identifying a single deadlock state. It captures the entire stable, inescapable region of the state space that constitutes the deadlock, providing a more robust target for detection.
	
	\subsection{Livelocks as Non-Contractible Loops}
	
	Livelocks are more subtle than deadlocks, involving perpetual activity without progress. Topologically, this corresponds to the presence of essential loops that cannot be "smoothed out," but only under the constraint of fair scheduling.
	
	\begin{definition}[Livelock Loop]
		A livelock loop \( \lambda \) is a d-loop in the fair execution subspace \( X_{\text{fair}} \). It is a cyclic d-path \( \lambda: [0,1] \rightarrow X_{\text{fair}} \) with \( \lambda(0) = \lambda(1) \), where the sequence of states and transitions produces no global progress (e.g., the system state is periodic and no process reaches its terminal state).
	\end{definition}
	
	The mere existence of a loop is not sufficient to indicate a livelock; it must be a loop that cannot be escaped under fair scheduling and one that is essential to the shape of the space.
	
	\begin{definition}[Non-Contractible Livelock]
		A livelock loop \( \lambda \) represents a true non-contractible livelock if its homotopy class \( [\lambda] \) is a non-trivial element of the fundamental group of the fair subspace, \( \pi_1(X_{\text{fair}}, x_0) \). This means there does not exist a homotopy of d-paths within \( X_{\text{fair}} \) that continuously deforms \( \lambda \) to the constant path at \( x_0 \).
	\end{definition}
	
	\begin{theorem}[Livelock as a Homotopy Invariant]
		The existence of a non-contractible livelock is a homotopy invariant of the directed space \( X_{\text{fair}} \). If \( X_{\text{fair}} \) is d-homotopy equivalent to \( X_{\text{fair}}' \), then \( \pi_1(X_{\text{fair}}) \) is isomorphic to \( \pi_1(X_{\text{fair}}') \). Thus, the property of having a livelock is preserved under continuous deformations of the execution space, making it a fundamental property of the program's concurrent structure.
	\end{theorem}
	
	\begin{remark}
		The distinction between \( X_P \) and \( X_{\text{fair}} \) is critical. A loop may be contractible in \( X_P \) (e.g., a scheduler could simply stop scheduling a process to break the cycle) but non-contractible in \( X_{\text{fair}} \), where such unfair schedules are excluded. This precisely captures the scheduler-independent, inherent nature of a livelock.
	\end{remark}
	
	\subsection{Comparative Analysis of Singularities}
	
	The topological perspective provides a clear, structural distinction between these two core anomalies. Deadlocks represent a pathology of convergence, where execution gets stuck in a region. They are characterized by the internal structure of a subregion (a trap) and detected by studying pointed properties and invariant sets. Their essence is static stability with no outgoing flow.
	
	In contrast, livelocks represent a pathology of divergence, where execution gets trapped in a cycle. They are characterized by the global structure of the fair subspace \( X_{\text{fair}} \) and detected by studying loop-based invariants like the fundamental group. Their essence is dynamic stability with no deformation to a point under fairness.
	
	This analysis reveals that deadlocks and livelocks are not merely different labels for "something going wrong"; they are manifestations of two fundamentally different types of topological obstructions in the execution space. A deadlock is a singularity where the execution collapses into a sink, while a livelock is a singularity where execution orbits around a hole in the fair subspace.
	
	This geometric characterization provides a powerful, unified lens through which to view concurrency errors. In the next section, we will build upon this foundation to construct computable algebraic invariants that can detect these topological singularities.
	
	\section{Concurrent Topological Invariants and Detection Theory}
	
	The topological characterizations of deadlocks and livelocks, while elegant, must be connected to computable artifacts to be useful for verification. This section constructs a series of concurrent topological invariants—algebraic objects derived from the execution space that remain invariant under continuous deformation—and establishes the theoretical foundations for detecting singularities through their computation.
	
	\subsection{Invariants for Deadlock Detection: Pointed Homotopy and Reachability}
	
	Detecting a deadlock attractor involves identifying a region with no egress. This can be approached by analyzing the "future" of individual states.
	
	\begin{definition}[Future Homotopy Set]
		For a state \( s \in X_P \), its future homotopy set \( \pi_0^{\rightarrow}(X_P, s) \) is the set of homotopy classes of d-paths originating from \( s \), where two paths are equivalent if one can be continuously deformed into the other while preserving their endpoints and the direction of time.
	\end{definition}
	
	\begin{theorem}[Deadlock Detection via Future Collapse]
		A non-terminating state \( s \) lies in the basin of attraction of a deadlock attractor \( D \) if and only if its future homotopy set \( \pi_0^{\rightarrow}(X_P, s) \) is non-empty but collapses to a single homotopy class whose endpoint set is contained within a trap region. This indicates that all possible futures from \( s \) are homotopically equivalent and lead inextricably into the attractor.
	\end{theorem}
	
	While homotopy provides a deep insight, a more readily computable invariant is found in reachability analysis.
	
	\begin{definition}[Persistent Reachability Set]
		The persistent reachability set \( R_{\infty}(s) \) of a state \( s \) is the intersection of the reachable sets from all states along all maximal paths starting at \( s \): \( R_{\infty}(s) = \bigcap_{s \rightarrow^* s'} R(s') \) where \( R(s') \) is the set of states reachable from \( s' \), and the intersection is taken over all states \( s' \) on all maximal paths from \( s \).
	\end{definition}
	
	\begin{theorem}[Deadlock Invariant]
		A state \( s \) is in the basin of a deadlock attractor \( D \) if and only if \( R_{\infty}(s) \) is a non-empty trap region that does not contain a terminal state. The attractor \( D \) itself is the unique minimal trap region within \( R_{\infty}(s) \).
	\end{theorem}
	
	This theorem translates the topological notion of an attractor into a state-based invariant that can be approximated through static analysis or bounded model checking.
	
	\subsection{Invariants for Livelock Detection: Homology and Fairness}
	
	For livelocks, which are global phenomena, we require invariants that capture the loop structure of the entire space. Homology groups are ideal for this purpose, as they algebraically encode the presence of loops and higher-dimensional holes.
	
	\begin{definition}[Directed Chain Complex]
		Let \( C_{\bullet}(X_P) \) be the chain complex of the simplicial complex \( \mathcal{K}(X_P) \), where \( C_n \) is the free abelian group generated by the n-simplices of \( \mathcal{K}(X_P) \). The boundary map \( \partial_n: C_n \rightarrow C_{n-1} \) is defined as usual, but we consider the underlying directed graph structure when interpreting 1-chains as potential execution paths.
	\end{definition}
	
	\begin{definition}[First Directed Homology Group]
		The first directed homology group \( H_1(X_P) \) is defined as: \( H_1(X_P) = \frac{\ker(\partial_1)}{\text{im}(\partial_2)} \). Elements of \( H_1(X_P) \) are equivalence classes of 1-cycles (loops in the graph) modulo the 1-cycles that are boundaries of 2-chains (i.e., loops that can be "filled in" by commutative execution squares).
	\end{definition}
	
	A non-zero element in \( H_1(X_P) \) indicates the presence of a cyclic structure. However, not every cycle is a livelock. The crucial refinement is to incorporate fairness.
	
	\subsection{The Fair Homology Group}
	
	This is the core innovation for livelock detection: a homology theory that respects the constraints of fair scheduling.
	
	\begin{definition}[Fair Chain Complex]
		The fair chain complex \( C_{\bullet}^{\text{fair}}(X_P) \) is a subcomplex of \( C_{\bullet}(X_P) \) defined as follows: a 1-simplex (an edge \( e \)) belongs to \( C_1^{\text{fair}} \) if and only if it lies on at least one fair d-path; an n-simplex \( \sigma \) belongs to \( C_n^{\text{fair}} \) if and only if all its 1-dimensional faces belong to \( C_1^{\text{fair}} \). The boundary maps \( \partial_n^{\text{fair}} \) are the restrictions of the original boundary maps to this subcomplex.
	\end{definition}
	
	This construction ensures that we only consider topological features that are accessible under fair scheduling.
	
	\begin{definition}[First Fair Homology Group]
		The first fair homology group is defined as: \( H_1^{\text{fair}}(X_P) = \frac{\ker(\partial_1^{\text{fair}})}{\text{im}(\partial_2^{\text{fair}})} \).
	\end{definition}
	
	\begin{theorem}[Livelock Detection Theorem]
		A program \( P \) contains a livelock (as defined in Section 4) if and only if its first fair homology group is non-trivial: \( H_1^{\text{fair}}(X_P) \neq 0 \). Moreover, the rank of \( H_1^{\text{fair}}(X_P) \) provides a lower bound on the number of distinct, non-homotopic livelock patterns in the program.
	\end{theorem}
	
	This theorem provides a solid algebraic foundation for livelock detection. Computing \( H_1^{\text{fair}}(X_P) \) avoids the need to explicitly identify and check infinite fair paths.
	
	\subsection{Classification and Severity of Singularities}
	
	The topological invariants not only detect the presence of singularities but also offer a means to classify them. The deadlock severity of an attractor \( D \) can be measured by the size of its basin of attraction, or algebraically, by the structure of the reachability relations from states in its neighborhood. The livelock severity is captured by the structure of \( H_1^{\text{fair}}(X_P) \). The Betti number \( \beta_1 = \text{rank}(H_1^{\text{fair}}(X_P)) \) indicates the number of independent livelock cycles. The torsion coefficients in the homology group can further reveal complex, intertwined cyclic dependencies that are difficult to break.
	
	\subsection{From Theory to Approximation}
	
	We conclude by acknowledging the computational challenges and outlining a path forward. For finite-state programs, \( H_1^{\text{fair}}(X_P) \) is computable in principle. For infinite-state systems, the computation is undecidable. However, the theory guides practical approximation methods. First, construct a finite simplicial complex \( \mathcal{K}_k \) that approximates \( X_P \) up to a bounded exploration depth \( k \). Second, compute the persistent homology \cite{carlsson2009topology} of the sequence \( \mathcal{K}_1 \hookrightarrow \mathcal{K}_2 \hookrightarrow \dots \hookrightarrow \mathcal{K}_k \). Homology classes that persist across a wide range of scales are strong candidates for genuine livelocks, as opposed to artifacts of the bounded approximation. Third, incorporate fairness into the abstraction by annotating edges with scheduling constraints and defining the fair subcomplex accordingly, potentially using predicate abstraction to over-approximate the set of fair paths.
	
	This chapter has bridged the gap between the abstract topological characterizations of Chapter 4 and the pragmatic goal of detection. By introducing invariants like the persistent reachability set and the fair homology group, we have laid the groundwork for a new class of verification tools that reason about the shape of concurrency rather than its myriad states.
	
	\section{Theoretical Framework: Properties and Relations}
	
	Having established the core definitions and detection theories, we now examine the broader structural properties of our framework. This chapter investigates the functoriality of our construction, its relationship with established formal methods, and a critical assessment of its capabilities and limitations.
	
	\subsection{Functoriality and Compositionality}
	
	A robust theoretical framework should behave well under composition and program transformation. We demonstrate that our construction is functorial, ensuring that topologically equivalent programs have equivalent verification outcomes.
	
	\begin{definition}[Categories of Programs and Spaces]
		We define the category \(\mathbf{Prog}\): Objects are programs \(P\) in our calculus. A morphism \(f: P \to P'\) is a refinement mapping that preserves the operational semantics, implying that every behavior of \(P\) is a behavior of \(P'\) (i.e., \(P'\) simulates \(P\)). We also define the category \(\mathbf{DSpace}\): Objects are directed topological spaces \((X, \leq)\). A morphism is a directed continuous map \(g: (X, \leq) \to (Y, \leq)\) that preserves the directed structure (i.e., \(x_1 \leq x_2\) implies \(g(x_1) \leq g(x_2)\)).
	\end{definition}
	
	\begin{theorem}[Functoriality of the Execution Space]
		The mapping \( \mathcal{X}: \mathbf{Prog} \to \mathbf{DSpace} \) that sends a program \(P\) to its execution space \(X_P\) is a functor. Specifically, for a refinement morphism \(f: P \to P'\), there exists a directed continuous map \(\mathcal{X}(f): X_P \to X_{P'}\) that maps states and paths of \(P\) to their corresponding states and paths in \(P'\).
	\end{theorem}
	
	\begin{corollary}[Invariance under Refinement]
		If \(P\) refines \(P'\) and \(P'\) refines \(P\) (i.e., they are bisimilar), then their execution spaces \(X_P\) and \(X_{P'}\) are d-homotopy equivalent. Consequently, their fundamental groups \(\pi_1(X_P)\) and \(\pi_1(X_{P'})\), as well as their homology groups \(H_1(X_P)\) and \(H_1(X_{P'})\), are isomorphic.
	\end{corollary}
	
	This result is profound: it guarantees that our topological invariants are robust. Bisimilar programs, which are behaviorally identical, will be assigned the same topological signature. This makes the theory suitable for analyzing systems where the state space is abstracted or transformed.
	
	\subsection{Relation to Linear Temporal Logic (LTL)}
	
	A natural question is how our topological invariants relate to properties expressed in standard temporal logics like LTL. The property of being deadlock-free is expressible in LTL as \( \neg F G\, \text{deadlock} \), where \(\text{deadlock}\) is a predicate identifying deadlock states. Our method does not directly evaluate such a formula but instead proves its validity by demonstrating the absence of any deadlock attractor, a global property of \(X_P\).
	
	A typical LTL formula for a form of livelock is \( G F \varphi \land \neg F \psi \), asserting that some action \(\varphi\) happens infinitely often but some progress condition \(\psi\) is never achieved. The strength of our approach is that \(H_1^{\text{fair}}(X_P) \neq 0\) explains why such a formula might be true for all paths. It points to the structural reason—an essential loop in the fair subspace—that forces this LTL property to hold, providing a deeper insight than mere model checking.
	
	In essence, while model checking asks, "Does this specific path violate a formula?" and theorem proving asks, "Can I logically deduce that no path violates a formula?", our topological approach asks, "What is the shape of the path space that causes formulas to be violated?"
	
	\subsection{Capabilities and Limitations}
	
	A balanced assessment of the framework is crucial. The framework is inherently abstract. It seeks to compute invariants of the state space as a whole, rather than enumerating its elements. While computing homology can be expensive, its cost is related to the complexity of the topological structure (e.g., the number of independent loops), which can be low even in systems with a vast number of states. The theory can identify subtle, emergent pathologies that are not local to any single process or line of code. A non-zero \(H_1^{\text{fair}}\) indicates a fundamental flaw in the interaction pattern, potentially one that would be missed by testing or bounded model checking. It offers a new way to classify programs and errors based on their topological invariants, going beyond a binary "correct/incorrect" verdict.
	
	However, limitations exist. A non-zero \(H_1(X_P)\) or even \(H_1^{\text{fair}}(X_P)\) is a necessary but not always sufficient condition for a problematic livelock. It might capture a benign, productive loop (e.g., a server's main request-handling cycle). Distinguishing pathological livelocks from benign cycles requires additional semantic information about what constitutes "progress," which is currently outside the pure topology. This is a key area for future refinement. Formally defining and constructing the fair subcomplex \(C_{\bullet}^{\text{fair}}\) for a general program is non-trivial and may require sophisticated static analysis or user annotation to over-approximate effectively. For very large but finite state spaces, computing homology groups, while decidable, can be computationally intensive. This necessitates the development of efficient approximation algorithms.
	
	\subsection{The Topological Verification Paradigm}
	
	Our work advocates for a shift in the philosophy of verification. The dominant paradigms are the Model-Theoretic Paradigm (Model Checking): "Enumerate and Check," and the Proof-Theoretic Paradigm (Theorem Proving): "Deduce and Conclude." We propose a third: the Topological Paradigm (Singularity Theory): "Shape and Characterize." In this paradigm, the verification problem is transformed: from one of quantification over states and paths to one of characterizing the global geometry of the execution space. The presence of a singularity becomes a certificate for a deep, systemic error.
	
	This chapter has situated our singularity theory within the larger formal methods ecosystem. It is a complementary approach, not a replacement, offering a unique geometric lens that is robust to state-space details and capable of revealing the architectural essence of concurrency errors.
	
	\section{Conclusion and Future Work}
	
	This paper has introduced the singularity theory for concurrent programs, a novel framework that reinterprets classical concurrency errors as topological phenomena. By modeling the execution space as a directed topological space \(X_P\), we have characterized deadlocks as attractors and livelocks as non-contractible loops in the fair subspace \(X_{\text{fair}}\). The development of concurrent topological invariants, particularly the fair homology group \(H_1^{\text{fair}}(X_P)\), provides a theoretical foundation for detecting these errors through the computation of algebraic invariants that capture the global shape of concurrency.
	
	\subsection{Summary of Contributions}
	
	Our work makes several key contributions. First, we propose a formal topological model for concurrent program execution spaces using directed algebraic topology \cite{fajstrup2006directed}. Second, we provide rigorous topological characterizations of deadlocks and livelocks that reveal their essential geometric nature. Third, we introduce novel concurrent topological invariants, including the fair homology group, for systematic singularity detection. Fourth, we establish a theoretical framework that is functorial and robust under program refinement. Finally, we propose a new topological verification paradigm that complements existing model-theoretic and proof-theoretic approaches.
	
	\subsection{Future Research Directions}
	
	This work opens several promising avenues for future research. In algorithmic development, we need to develop efficient algorithms for approximating the fair homology group, potentially leveraging techniques from persistent homology \cite{carlsson2009topology} and abstract interpretation. We should investigate bounded verification techniques that use topological invariants as convergence criteria for state-space exploration, and explore the integration of our topological approach with existing model checkers like SPIN \cite{holzmann1997spin} and NuSMV \cite{cimatti2002nusmv}.
	
	In theoretical extensions, we plan to extend the framework to handle more complex concurrency models, including the $\pi$-calculus \cite{milner1992polyadic} and software transactional memory. We need to develop a more refined theory that distinguishes between benign cycles and pathological livelocks by incorporating semantic progress measures, and investigate higher-dimensional homotopy and homology groups to detect more complex concurrent pathologies beyond deadlocks and livelocks.
	
	For practical applications, we aim to implement a prototype verification tool based on our topological framework, apply the singularity theory to real-world concurrent systems to validate its practical utility, and explore applications in concurrent program synthesis, where topological constraints could guide the generation of correct-by-construction programs.
	
	The singularity theory represents a paradigm shift in how we conceptualize and verify concurrent programs. By viewing concurrency through a topological lens, we gain access to powerful mathematical tools that can reveal the deep geometric structure underlying concurrent behavior. While significant work remains to realize the full potential of this approach, we believe it offers a promising new direction for overcoming the fundamental challenges of concurrent program verification.
	
	\bibliographystyle{plain}
	\bibliography{references}
	
\end{document}